\documentclass[twocolumn,amssymb,amsmath,aps,prb]{revtex4-2}
\usepackage{graphicx}
\usepackage{bm}
\begin{document}
\title{Magnetization, dielectric and thermal studies in the double perovskite polycrystalline  compound Tm$_2$CoMnO$_6$}
\author{A. Banerjee$^{1,2}$}
\email{anupam697@gmail.com}
\author{Gangadhar Das$^{3,4}$}
\author{V. Rajaji$^{3,5}$} 
\author{S. Majumdar$^2$}
\author{P.K. Chakrabarti$^1$}
\affiliation{$^1$Solid State Research Laboratory, Department of Physics, Burdwan University, Burdwan 713104, India}
\affiliation{$^2$School of Physical Sciences, Indian Association for the Cultivation of Science, 2A \& B Raja S. C. Mullick Road, Jadavpur, Kolkata 700032, India}
\affiliation{$^3$Chemistry and Physics of Materials Unit, Jawaharlal Nehru Centre for Advanced Scientific Research, Jakkur, Bangalore 560 064, India}
\affiliation{$^4$Elettra Sincrotrone Trieste, Strada Statale 14, km 163.5 in AREA Science Park, Basovizza, Trieste 34149, Italy}
\affiliation{$^5$University Lyon, Université Claude Bernard Lyon 1, CNRS, Institut Lumière Matière, F-69622 Villeurbanne, France}

\begin{abstract} 
We report here a comprehensive study on structural, magnetic, caloric and electronic properties of the monoclinic phase of double perovskite compound Tm$_2$CoMnO$_6$ (TCMO) in its polycrysttaline format. Magnetic measurements confirm the presence of thermal hysteresis in magnetization indicates towards a first order magnetic transition at its critical point ($T_c$). Our study finds the metamagnetic jump in isothermal $M(H)$ at 2 K that signifies the rare earth spin allignement in loop cycling above a certain critical field. This jump is only present at very low $T$ that dies out as the $T$ raises. $M(H)$ measurements at different temperature stipulate that Co$^{2+}$ and Mn$^{4+}$ do not order completely inspite of strong ferromagnetic correlation that is against the previous study on TCMO. The concurrence of structral, magnetic and dielectric anomaly at $T_c$ suggest its possible magnetostructural copuling. We found a high dielectric constant ($\sim$ 2700) in TCMO and the Maxwell-Wagner electrial loss spectrum analysis hints towards a long range hopping mechanism. Sizeable magnetocaloric effect in terms of entropy change across the transition temperatures has been obtained from heat capacity data that clearly corroborates its intrinsic behaviour. 
\end{abstract} 
\maketitle

\section{Introduction}
From the last century, scientists have been ceaselessly exploring magnetic materials, especially magnetic oxides to find out various exotic functional properties and to utilize them in wide range of technological applications such as memory devices, medical techniques, energy storage, CFC-free refrigeration, remote sensing devices and many more ~\cite{memory1,memory2,medical,energy1,energy2,remote}. Double perovskite (DP) oxides having general formula A$_2$BB$^{\prime}$O$_6$, comprising of magnetic ions are such a system where several unusual and interesting features like colossal magneto-resistance, multiferroicity ~\cite{sharma,chikara,terada,kim}, large magnetocaloric effect ~\cite{moon1,moon2,balli}, spin phonon coupling ~\cite{liu}, exchange bias~\cite{liu,sfco,murthy}, metamagnetism ~\cite{metamag1,cao,kim2,ding} have been reported. Particularly, DP manganites having rare earth (R) ions at A-site have been the focus of research as manganese offers stable mixed valency in some compounds leading to charge localization, phase separation, glassy behaviour etc. Moreover, rare earth ions enrich magneto-crystalline anisotropy and spin-orbit coupling. For example, DP manganites with Gd and Tb ions, the spiral spin modulation can induce ferroelectricity through antisymmetric exchange striction whereas an improper ferroelectricity due to structural trimerization can be observed in case of smaller R-ions (such as Er or Tm) ~\cite{kimura}.
\par
R$_2$CoMnO$_6$ (R = rare earth metals La-Lu, and Y) (RCMO) is one of the well explored DP manganites as they exhibits diverse magnetic and functional properties. From the neutron diffraction studies, it is evident that  there are two types of spin orderings, namely, (i) E-type: characterized by up-up-down-down ($\uparrow\uparrow\downarrow\downarrow$) arrangements of Co$^{2+}$ and Mn$^{4+}$ spins along a particular crystallographic direction resulting competing nearest neighbour ferromagnetic(FM) and next-nearest neighbour antiferromagnetic (AFM) exchange interactions; and  (ii) F-type: having Co$^{2+}$ and Mn$^{4+}$ spins in FM collinear arrangements ~\cite{tmcmo,kumar,veni,tapan_chatt,kim3}. RCMO with smaller size R-ions, such as Lu$_2$CoMnO$_6$, Y$_2$CoMnO$_6$, Yb$_2$CoMnO$_6$ have E-phase and they show ferroelectricity that emerges perpendicular to the $c$-axis through the symmetric exchange striction ~\cite{sharma,kim,veni}. For Yb$_2$CoMnO$_6$, on application of a  magnetic field, E-type phase destabilizes to F-type through a metamagnetic jump. As the size of the R-ions decreases, the magnetic transition temperature, arising from the dominant Co$^{2+}$ and Mn$^{4+}$ superexchange interactions, decreases monotonously from 204 K (La$_2$CoMnO$_6$) to 48 K (Lu$_2$CoMnO$_6$) ~\cite{kim3}. It becomes more exotic when R-ions start to show long range magnetic ordering. In Er$_2$CoMnO$_6$, Er$^{3+}$ moments align opposite to FM Co$^{2+}$/Mn$^{4+}$ sublattices activates ferrimagnetic ordering ~\cite{banerjee} below about 30 K. Moreover, metamagnetism and strong anistropy of Er$^{3+}$ moments generates an unusual inverse hysteresis loop. Occasionally, stacking faults give rise to antisite disorder (ASD) that causes exchange bias in RCMO compounds ~\cite{metamag1,disorder}. 
\par 
Here, we studied DP compound Tm$_2$CoMnO$_6$ (TCMO) through structural, magnetic, thermal and dielectric analyses. Earlier works on this compound revealed its magnetic structure, where Tm$^{3+}$ moments coupled antiferromagnetically to F-type arrangements of Co$^{2+}$/Mn$^{4+}$ spins ~\cite{tmcmo}. It was claimed and showed that an F-type arrangements didn't exhibit any dielectric anomaly originated from the magnetic order. However, some of RCMO with larger R-ions show dielectric anomalies at the magnetic transition point, which renders further clarification. TCMO has large magnetic moment due to the presence of magnetic elements Tm, Co and Mn, and the magnetic exchange interaction is supposed to be complex. The observation of metamagnetism provides another important evidence for the magnetic instability. It is therefore, essential to carry out in-depth magnetic study of the sample in the light of its magnetic structure provided by the previous neutron diffraction study. The large moment of the sample along with metamagnetism  can also be imperative for showing magneto-functional propeties such as magneto-caloric effect.

\begin{figure}[t]
\centering
\includegraphics[width = 9 cm]{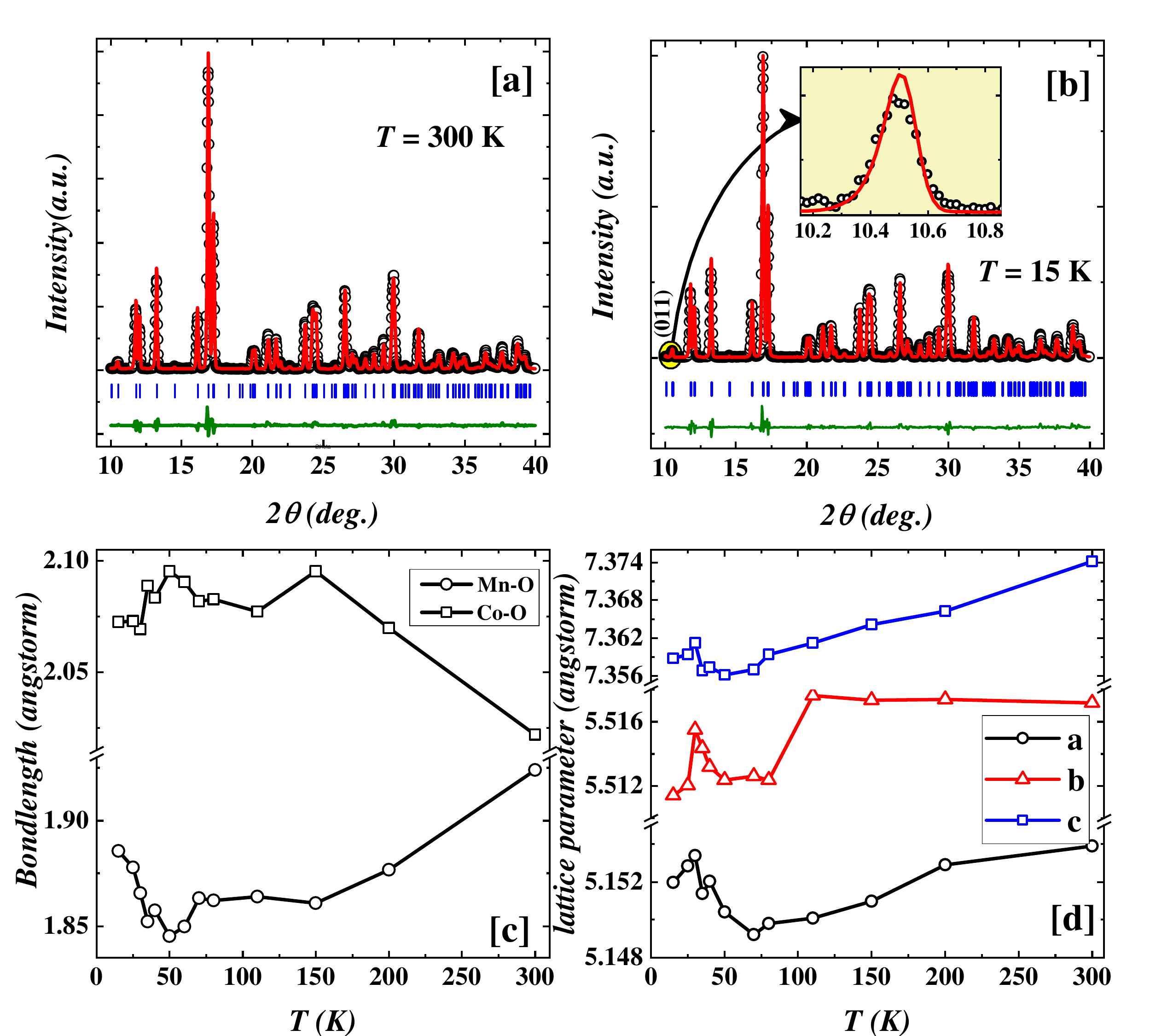}
\caption {(Color online) (a) and (b) show the X-ray diffraction patterns of Tm$_2$CoMnO$_6$ recorded at 300 and 15 K respectively. Observed (black open circles), calculated  (continuous red line), difference (continuous green line) intensities and allowed Bragg positions(blue ticks) are marked in the figure. (c) depicts the thermal variation of Co-O and Mn-O bond distance respectively. (d) displays the temperature dependence of the lattice parameters.}
\end{figure}

\section{Experimental details}
The synthesis of the bulk polycrystalline sample of Tm$_2$CoMnO$_6$ was carried out through solid state reaction route. The powders of Tm$_2$O$_3$, CoO and MnCO$_3$ were intimately mixed in proper stoichiometric ratio and heated at 1423 K  for 24 hours. Thereafter, the mixture was pressed into pellets, and they were annealed in air for 3 days at 1623 K with several intermediate grindings. The sample was investigated by powder x-ray diffraction (PXRD) measurement using Cu K$_{\alpha}$ radiation and the pattern confirm its reported structure with phase purity. Further advanced structural studies were done through high resolution temperature dependent PXRD (wavelength of the radiation being 0.855 \AA) using synchrotron facility at photon factory, KEK, Japan at different temperatures ($T$) ranging between 15 to 300 K. 
\par
The magnetic measurements were performed on a Quantum Design SQUID magnetometer (MPMS Evercool model) as well as on a vibrating sample magnetometer (VSM). Before each low-$T$ measurement, the sample was heated above critical temperature, and the magnet was degaussed. Heat capacity ($C_p$) data, were recorded using a commercial Quantum Design physical properties measurement system (PPMS, Dynacool model). Dielectric measurements have been carried out using a He-cryocooler (Advanced Research Systems, Inc), LCR meter (WayneKerr; model-4100) and a temperature controller (LakeShore; model-331).

\begin{figure}[t]
\centering
\includegraphics[width = 9 cm]{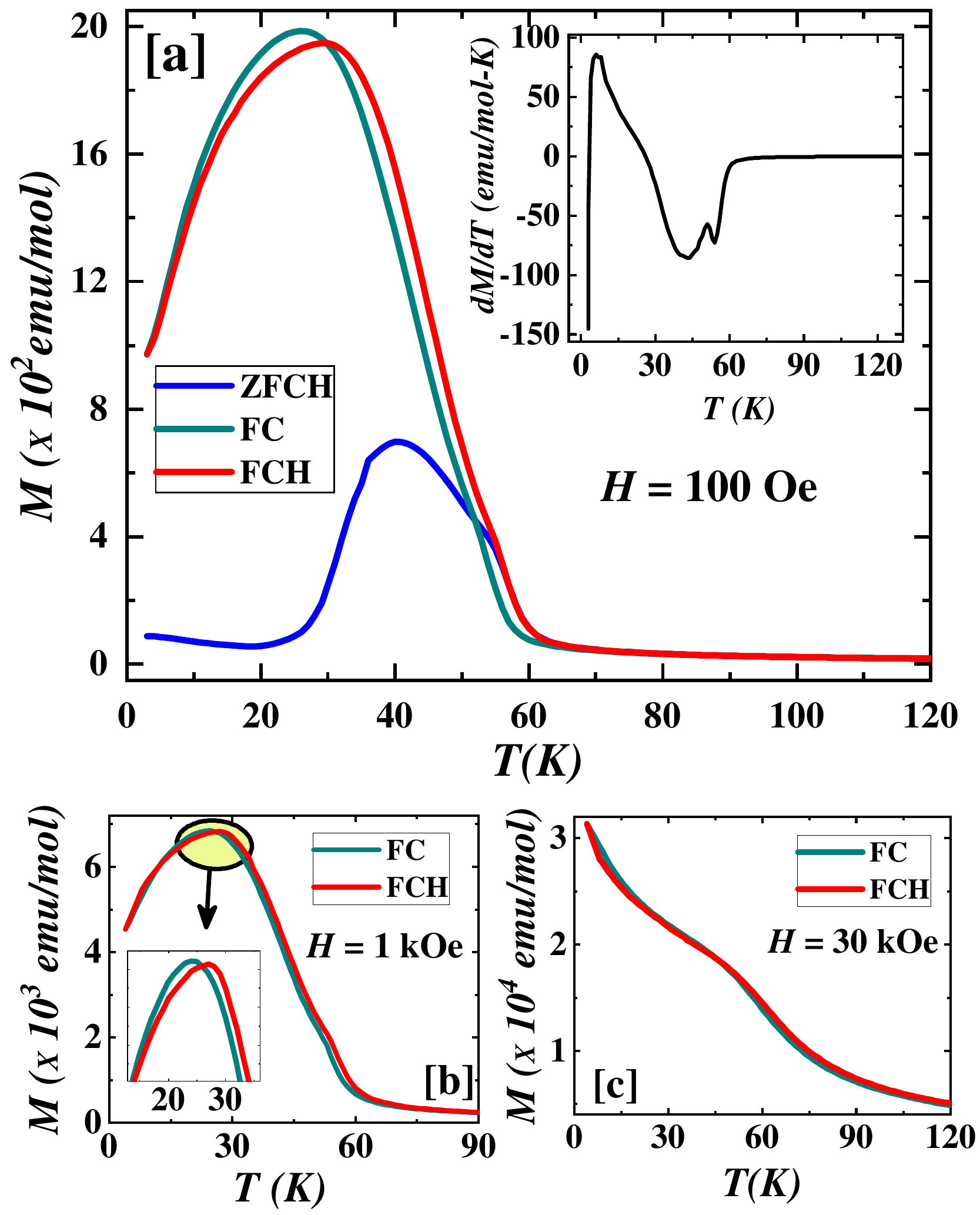}
\caption{(Color online) (a) show magnetization as a function of temperature measured in zero-field-cooled-heating (ZFCH), field-cooling (FC) and field-cooled-heating (FCH) protocols with 100 Oe magnetic field whereas the inset shows thermal variation of $\frac{dM}{dT}$ of FC data. (b) and (c) plot the M(T) data in FC and FCH mode for $H$= 1 kOe and 30 kOe respectively. The inset of (b) zooms in the circled portion.}
\end{figure}

\begin{figure}[t]
\centering
\includegraphics[width = 9 cm]{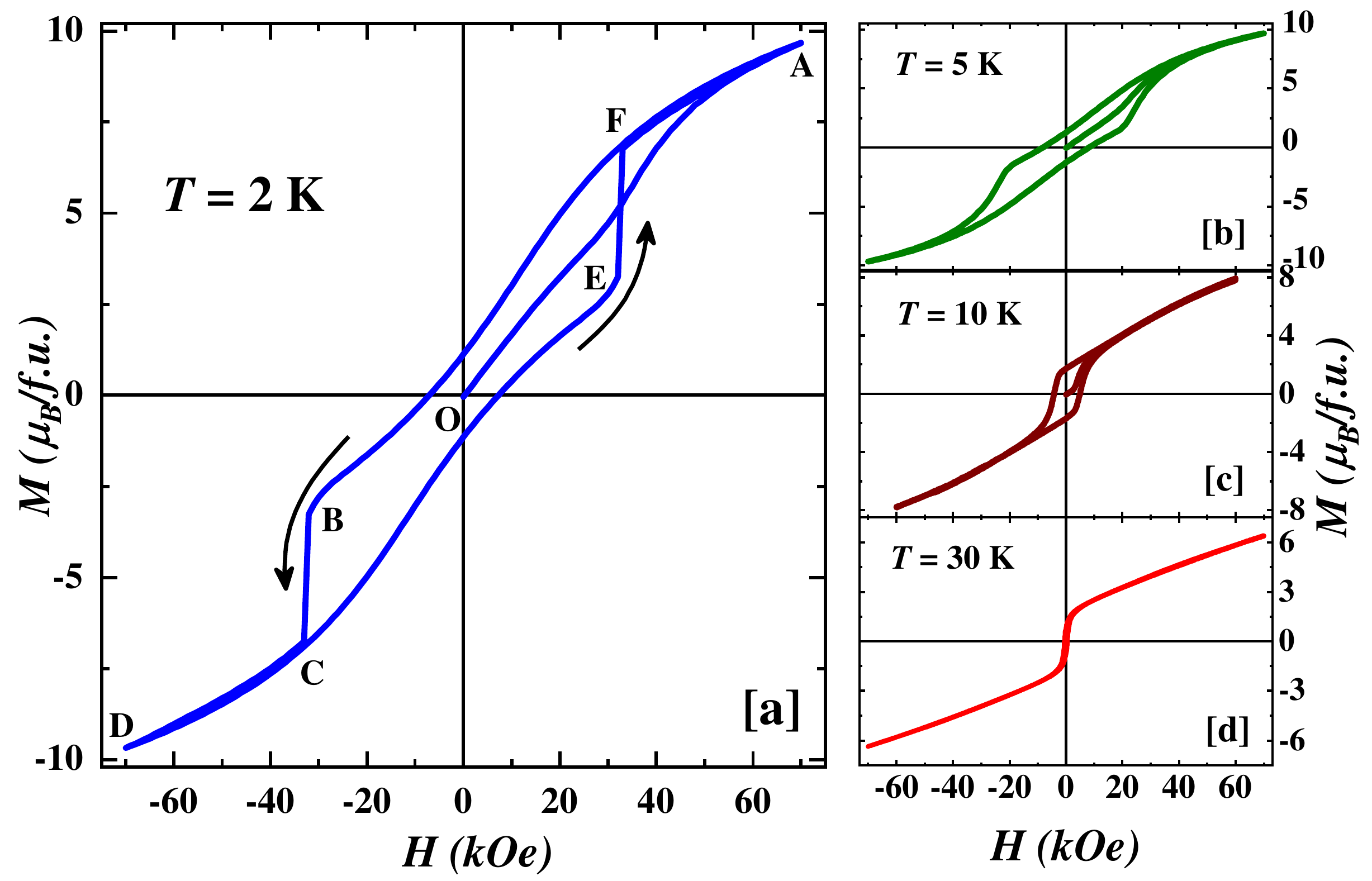}
\caption{(Color online) (a)-(d) display the isothermal magnetization data measured at 2, 5, 10 and 30 K respectively in the ZFC condition.}
\end{figure}
\section{Results}

\subsection{Crystal structure} 
 We have performed Rietveld refinement of the synchrotron data using MAUD suit~\cite{maud} and two representative fitted data (for 300 K and 15 K) are shown in figs. 1 (a) and (b) respectively. The experimental data converges well with the refined pattern having a monoclinic double perovskite structure (space group $P2_1/n$) with Co$^{2+}$ and Mn$^{4+}$ ions occupying alternating corner-shared CoO$_6$/MnO$_6$ octahedra. For the perfectly ordered structure, one would expect (011) superlattice reflection for the present P2$_1$/n structure. In our data, we see clear presence of the (011) line [shown in the inset of fig. 1(b)] indicates a good degree of ordered arrangement of Co and Mn. However, it is difficult to comment about the degree of order in the system quatitatively from the present measurement, and there can still be some antisite disorder between Co and Mn. This is due to that fact that DP are well-known to show ASD between B and B$^{\prime}$ sites~\cite{disorder,cation_disorder1,cation_disorder2,cation_disorder3}. The temperature variations of the bond lengths between Co-O and Mn-O atoms are plotted in fig. 1(c), which shows a non-monotonous behavior. The thermal variation pattern of Co-O shows a dip around the magnetic transition, while the curve for Mn-O depicts a peak at similar temperatures. However, we do not observe any change in the lattice symmetry associated with this structural anomaly. On the other hand, we observe anomalies in the temperature variation of different lattice parameters ($a$, $b$ and $c$). A broad dip like feature is present at around 60 K and a peak near at 27 K which will be known as magnetic transition temperatures ($T_c$ and $T_{N}^{Tm}$ respectively) in the following section. 
 
\subsection{Magnetization}
The thermal variation of magnetization ($M$) was measured under different applied magnetic fields ($H$) in the zero-field-cooled heating (ZFCH), field-cooling (FC) and field-cooled-heating (FCH) conditions. Fig. 2(a) shows such $M (T)$ curves with $H$ = 100 Oe in three different protocols, where an upturn below 60 K ($T_c$) is noticed. The transition temperatures can be better viewed in the curve of $T$ variation of the first order temperature derivative of magnetization ($\frac{dM}{dT}$) for FC data as shown in the inset of fig. 2(a). It shows a minimum around 43 K ($T_p$) with an additional kink at 55 K which is also identifiable in the data from main panel of fig. 2(a). The ZFCH data show a peak around $T_p$ and separate out from the FC and FCH data below the peak. The sharp rise below $T_c$ in FC and FCH magnetization data is likely to be associated with the ferromagnetic order of Co and Mn ions ~\cite{tmcmo,banerjee}. On further lowering the temperature, both FC and FCH data attain maximum at $T_{N}^{Tm}$ (= 27 K for FC data). At relatively lower temperature below $T_{N}^{Tm}$, $M$ drops monotonously as the Tm moments start to order in opposite way to the orientation of Co and Mn moments. Additionally, we can notice a sharp peak in $\frac{dM}{dT}$ vs $T$ plot near about 8 K ($T_s$) that can be assigned to the spin orientation of Tm-moments ~\cite{Gd-order}. There is a strong divergence between the ZFCH and FC data for $H$ = 100 Oe. These data are consistent with the previous magnetic measurement results.~\cite{tmcmo} $M(T)$ shows a well behaved Curie-Weiss nature down to 100 K. The effective moment is found to be 11.65 $\mu_B$/f.u., that corresponds to the expected moment from Tm$^{3+}$ (total angular momentum, $J =$ 6), Co$^{2+}$ (high spin, $S =$ 3/2) and Mn$^{4+}$ ($S =$ 3/2) ions in the sample.~\cite{tapan_chatt,kim3,tmcmo} The paramagnetic Curie temperature is found to be 23 K indicating the existence of a predominant ferromagnetic correlations between the spins. 
\par
It can be noted that the peak positions at $T_{N}^{Tm}$ for FC and FCH data are separated by around 4 K. Indicatively, we notice the presence of thermal hysteresis between FC and FCH data below 60 K to the lowest measured point. This signifies to the weak first order nature of the magnetic transition at $T_c$ ~\cite{banerjee2,siruguri}. The thermal irreversibilty between the FC and FCH data is still present at applied field $H$ = 1 kOe [fig. 2(b)], but the amount of hysteresis decreases as the applied field is higher. Notably, the magnetization pattern appears markedly different for $H$ = 30 kOe as depicted in fig. 2(c), where the maximum at $T_{N}^{Tm}$ is completely absent. 
\begin{figure}[t]
\centering
\includegraphics[width = 9 cm]{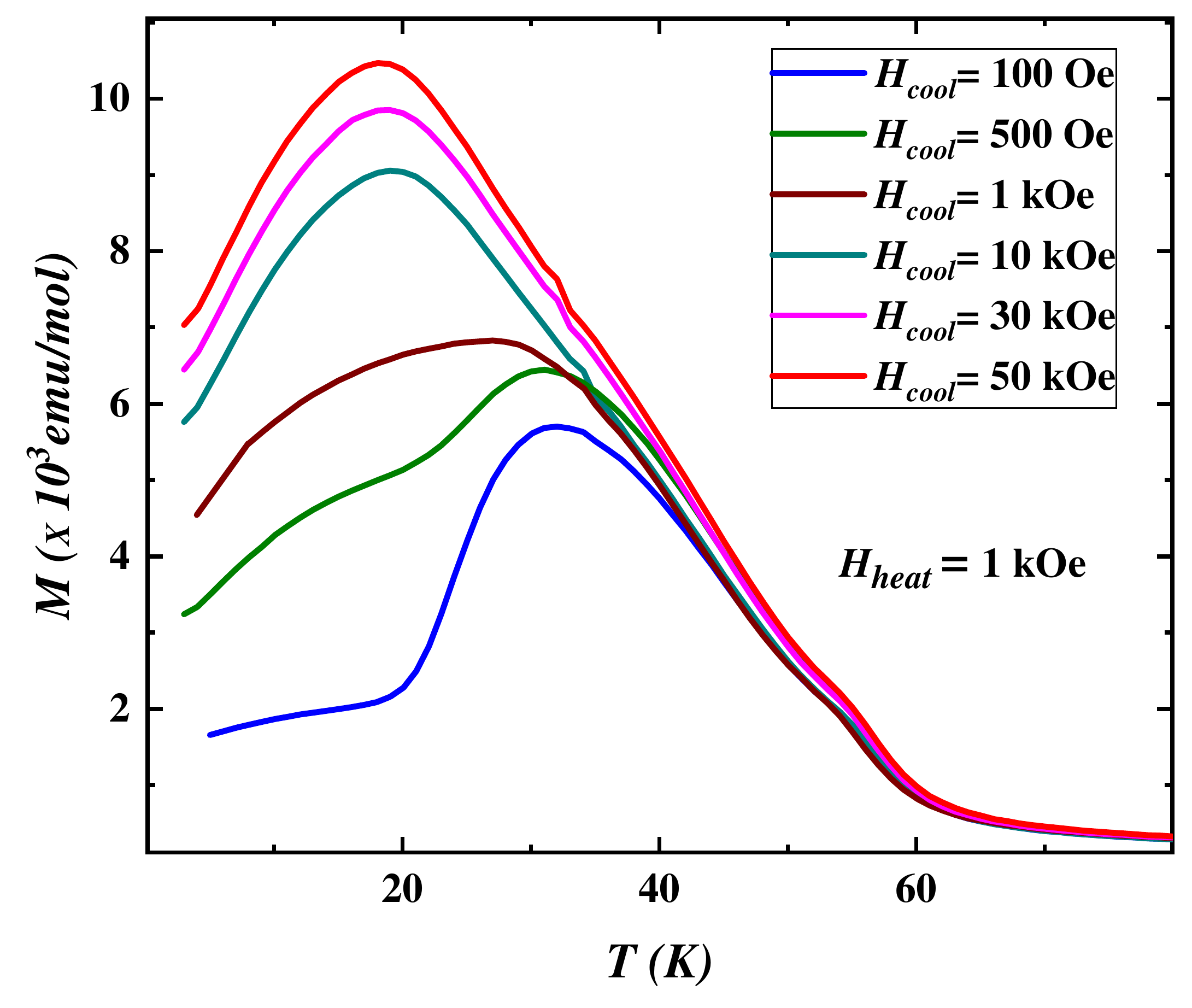}
\caption{(Color online) shows the cooling field dependence of magnetization ($M$), where M is measured in 1 kOe of field while heating after being cooled in $H_{cool}$ = 0.1, 0.5, 1, 10, 30 and 50 kOe from more than 100 K.}
\end{figure}
\par
Fig. 3(a) displays the isothermal magnetization curve as a function of $H$  at 2 K. The curve shows large hysteresis around the origin with a significant amount of coercive field, $H_{coer} =$ 7.2 kOe and remanent magnetization of 1.12 $\mu_B$/f.u. The $M$ does not show any kind of saturating tendency even at $H$= 70 kOe. Existence of non-saturated $M$ as well as the sharp drop in $M(T)$ at low temperature excludes a pure ferromagnetic-like state configuration ~\cite{kim3,banerjee2}. The amount of hysteresis in $M(H)$ decreases as we increase the temperature and vanishes above 30 K. The most interesting characteristic is the sharp metamagnetic jump at $H_c$ = $\pm$32 kOe in the field increasing leg except the virgin curve ~\cite{banerjee,banerjee2}. As we increase the temperature to 5 K [fig. 3(b)], the sharp jump tends to become broader, and the critical field for transition ($H_c$) lowers to almost 20 kOe. As we further increase temperature the signature of metamagnetism dies out [figs. 3(c-d)]. 

\par
TCMO shows narrow thermal hysteresis in its heating and cooling curves in $M(T)$ data [see fig. 2(a)], and the $T$-driven magnetic transition appears to be weakly first order in nature. In addition, the sample shows sharp metamagnetic transition along with field hysteresis indicating $H$ driven first order transition. Therefore, it is important to study the possible field-induced metastability across the first order transition. It has been already known that on field cooling across a first order magneto-structural transition, sample can retain its high-field state even when the field is withdrawn. Such kinetically arrested state can be highly metastable and it can disappear on external perturbation, such as, temperature or pressure. An elegant way to probe such metastability is provided by measuring magnetization while heating in a specific field after being field-cooled in different fields ($H_{cool}$) ~\cite{banerjee2,sbroy,chadda}. In fig. 4, we measured $M$ as a function of $T$ under $H$= 1 kOe of field while heating after cooling in different $H_{cool}$ (= 0.1, 0.5, 1, 10, 30 and 50 kOe.)s. It is clear from the graph that the curves for different $H_{cool}$ differ markedly from each other below about 40 K and they merge completely at 60 K, which is $T_c$ as stated in fig. 2(a). Additionally, the nature of the heating curves are different as $H_{cool}$ varies. In $H_{cool}$ = 100 Oe curve, there is a prominent change in slope around 25 K, which is defined as $T_{N}^{Tm}$ of the sample. Such change in slope disappears slowly with increasing $H_{cool}$ and it is completely absent for $H_{cool}\geq$ 10 kOe. The curves recorded at $H_{cool}$ = 10, 30 and 50 kOe are found to be separated from the other low field curves and they show rather uncomplicated feature with a large peak just below 20 K. Similar peaks are also present in the low field data, but they occur at higher temperatures for lower values of $H_{cool}$.   
\begin{figure}[t]
\centering
\includegraphics[width = 9 cm]{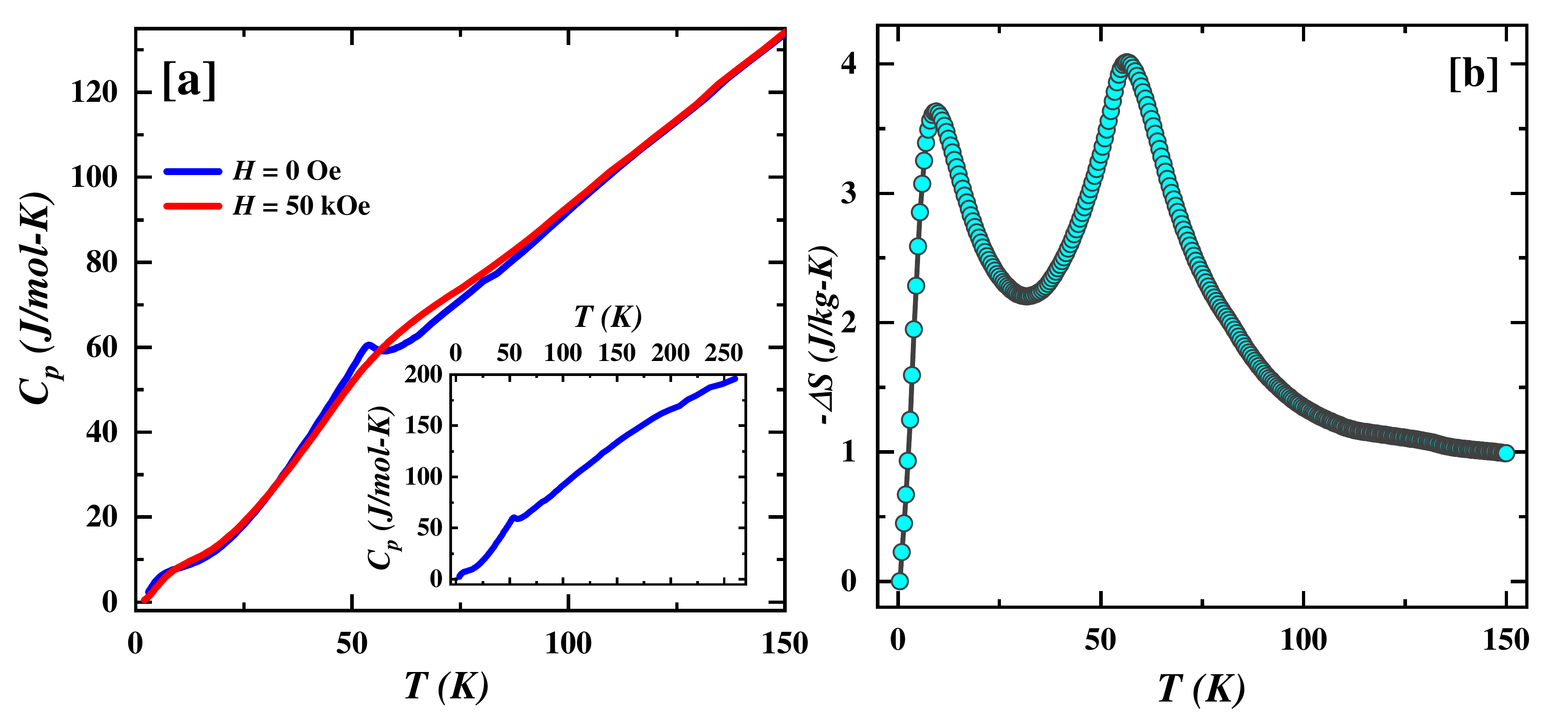}
\caption{(Color online) (a) shows both the zero field and 50 kOe heat capacity ($C_p$) as a function of temperature. Inset shows the zero field $C_p$ data over the complete measured temperature range, while (d) shows the thermal variation of the adiabatic change in entropy due to the application of H = 50 kOe.}
\end{figure}
\subsection {Heat Capacity}
To elucidate the field and temperature driven magnetic transition, we measured the heat capacity ($C_p$) of the sample at $H$ = 0 and 50 kOe [Fig. 5(a)]. A clear anomaly is seen at around $T$ = 60 K, which is matching with the $T_c$ obtained from our $M(T)$ data. There is an additional broad hump at low-$T$ around 8 K that can be corroborated with the spin orientation of Tm-moments ($T_s$) as observed in our magnetization data as well as in other double perovskites reported in the literature~\cite{kim,Gd-order,nag}. The 50 kOe $C_p$ data have been found to differ [Fig. 5, main panel] from the zero field curve below about 100 K and the peak observed around $T_c$ gets suppressed under the application of field. The field data join with the zero field one below about 40 K, however, a further deviation is seen at around the low-$T$ hump ($\sim$ 10 K).  
\par
We have investigated the magneto-caloric effect (MCE) in terms of the change in entropy $\Delta S$ due to the application of 50 kOe of magnetic field. Estimation of $\Delta S$ is performed through the following relation $$\Delta{S}(0\rightarrow H_0) = \int^{H_0}_{0}\left(\frac{C_p(T,H)-C_p(T,0)}{T}\right )dT,$$ where $\Delta{S}(0\rightarrow H_0)$ denotes the entropy change for the variation in $H$ from 0 to $H_0$ ~\cite{mce}. The temperature profile of $-\Delta S(T)$ is plotted in Fig. 5(b), where two peaks are observed at the similar transition points as $C_p(T)$ around $T_c$ = 60 K and $T_s$ = 8 K, which is associated with the suppression of spin fluctuations around the critical point by $H$. The peak value of the anomaly at $T_c$ is 4 J.kg$^{-1}$K$^{-1}$ and at the low-$T$ peak around 8 K, with the peak value of 3.6 J.kg$^{-1}$K$^{-1}$.

\begin{figure}[t]
\centering
\includegraphics[width = 9 cm]{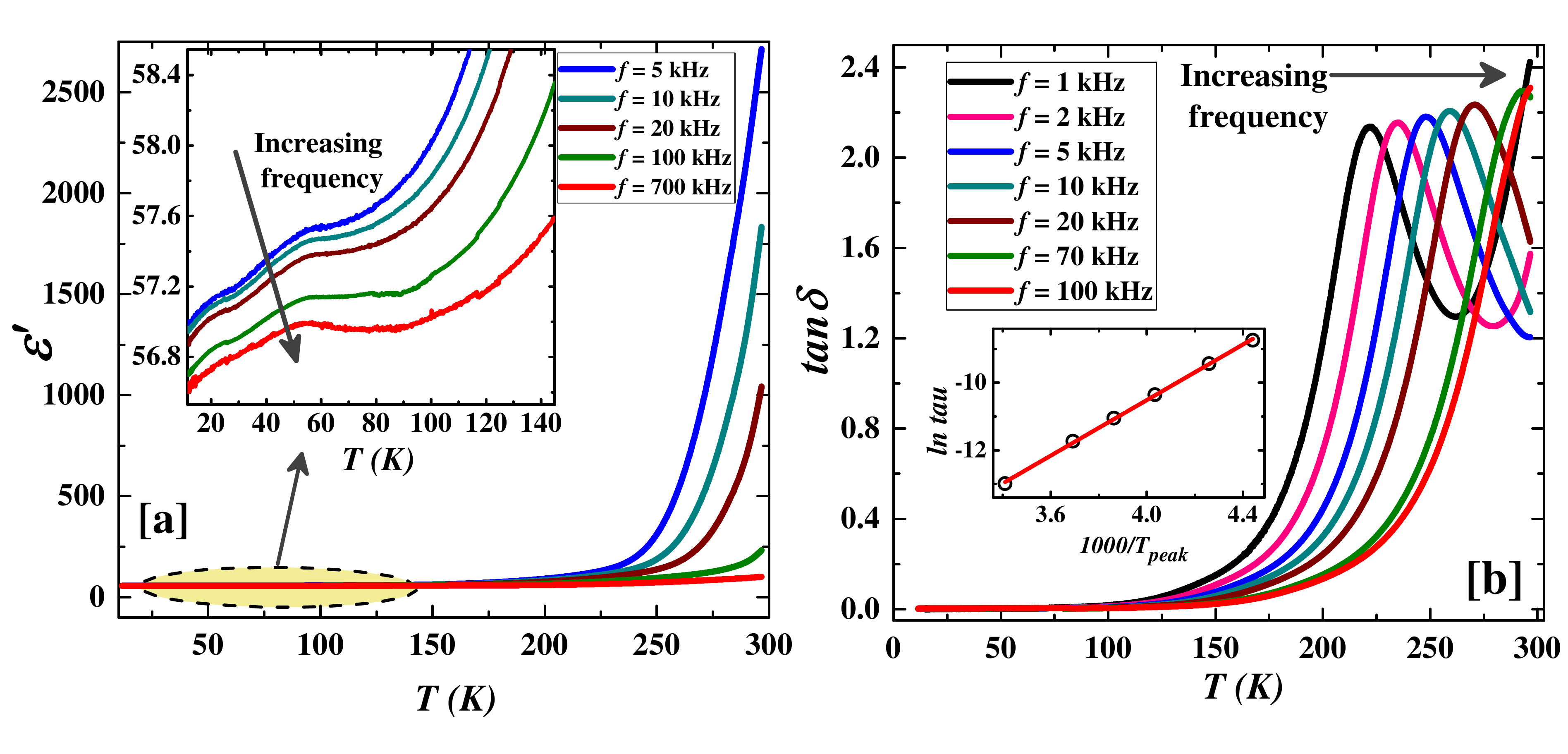}
\caption{(Color online) (a) shows the thermal variation of the real part of dielectric permittivity for various applied frequencies whereas inset magnifies the region of anomaly. (b) plots the loss tangent as a function of temperature for different frequencies. The inset displays $ln \tau$ versus 1000/$T_{max}$ plot where $T_{max}$ is obtained from the $\tan \delta$ versus $T$ plot by noting the peak positions. The solid line is the Arrhenius-type fitting to the data.}
\end{figure} 

\subsection {Dielectric studies}
All of the compounds in R$_2$CoMnO$_6$ series show insulating behaviour ~\cite{transport,wang} and so as Tm$_2$CoMnO$_6$. Our transport measurement indicate (not shown here) that the resistivity increases  with lowering of $T$ from 300 K. Here, we have studied thermal dependency of complex dielectric permittivity of Tm$_2$CoMnO$_6$ down to 10 K in zero field at different frequencies ($f$). The real part of the dielectric permittivity ($\epsilon^{\prime}$) has a considerably large value of $\sim$ 2700 at $f$ =5 kHz and room temperature [fig. 6(a)]. This type of compounds are considered to be high dielectric materials ~\cite{highEpsilon,banerjee3}. Interestingly, a hump like anomaly is observed at 53 K that is shown in the inset of fig. 6(a). The contiguity of transition temperature to the onset of magnetic ordering and the non-dispersive (in $f$) nature of the hump-like feature indicate to the existence of electrical order. The low-$T$ values of $\epsilon^{\prime}$ are almost $f$ independent, and it display its static value  originating  from the intrinsic contribution. Thereafter, it rises sharply as the $T$ increases. This step-like anomaly shifts to higher $T$ with increasing $f$ that describes its strong dispersion over $f$ range. This is associated with the peaks in $T$ variation of loss tangent ($\tan \delta = \epsilon^{\prime\prime}/\epsilon^{\prime}$) displayed in fig. 6(b), which show large frequency dispersion. The peak in $\tan{\delta(T)}$ describes an electrical loss, and such $f$-dependent peak in $\tan{\delta(T)}$, which is usually attributed to the occurrence of Maxwell-Wagner relaxations connected to the interface of grain boundary and electrodes as reported earlier ~\cite{tmcmo,wang,yang,lunk}. The peak temperature ($T_{peak}$) in $\tan \delta(T)$ differs appreciably with the mid point of the step-like increment in $\epsilon^{\prime}(T)$ and it is an evidence for the absence of permanent dipole moment in our system at least above 100 K. For the lower $f$ measurement, $\tan \delta$ rises again after the peak, pointing towards a strong electronic contribution at higher $T$. Furthermore, we have analyzed the $T_{peak}$ and find out the characteristic relaxation time from the relation $\tau = \frac{1}{2\pi f}$. The values of $\tau$ is plotted against $1000/T_{peak}$ in the inset of fig. 6(b) and it varies linearly. Evidently, it follows a Arrhenius type behaviour accordingly with $\tau = \tau_0 \exp(\frac{E}{k_BT})$, where $E$ is the activation energy and $\tau_0$ is the average relaxation time. From the fitting of $ln \tau$ vs $1000/T_{peak}$, we have obtained $E = 354.5$ meV and $\tau_0 = 1.94 \times 10^{-12}$ s. The value of $\tau_0$ is too large for normal electronic conduction ($\sim 10^{-16}$ s)~\cite{bid}, and therefore indicates towards a different charge carrying mechanism. The activation energy is also a bit higher, which signifies some kind of long range hopping in the system ~\cite{banerjee3}.

\section{Discussion}
Our investigation on Tm$_2$CoMnO$_6$ has revealed several important phenomena such as (i) the thermal hysteresis between FC and FCH data in low-field $M(T)$ pattern, (ii) metamagnetic jump at isothermal $M(H)$ curve at low $T$ (iii) existence of magnetic field induced arrested metastable states (iv) anomalous caloric effect at transition temperatures and (v) occurrence of dielectric anomaly at the magnetic transition temperature. 
\par
The observed thermal hysteresis in the $M$ versus $T$ heating and cooling data [FCH and FC respectively in Fig. 2(a)] indicates first order phase transition. The observed hysteresis is quite broad spanning over a wide range of temperature. The hysteresis starts from $T_c$ and the FCH and FC data cross with each other around 30 K. Below 30 K, the FCH data lie below the FC data. Due this crossover, we effectively get two separate thermal hysteresis loops, and they may correspond to two different magnetic transitions. Similar crossing of FC and FCH data were observed in case of well known multiferroic CaMn$_7$O$_{12}$ sample ~\cite{camn7o12}. It is already clear from our $M$ and $C_p$ data that the sample shows multiple magnetic transitions. 
\par
A first order transition is often associated with structural anomalies ~\cite{siruguri,fsma,fsma2}. From our $T$ dependent PXRD data, we observe multiple anomalies in the lattice parameters. We observe a dip like feature in the lattice parameters [Fig. 1(d)] around the magnetic critical point $T_c$. There are relatively sharp peaks in $a$, $b$ and $c$ around 27 K, which match well with $T_{N}^{Tm}$ of the sample. Interestingly, we found an anomalous variation in Mn-O and Co-O bond lengths at $T_c$. This indicates that the magnetic transition is associated with structural changes, although the crystal symmetry remains unchanged. The existence of significant structural deformation and thermal hysteresis indicates to the possible magneto-structural coupling that corroborates nicely to the magneto-crystalline anisotropy in the system ~\cite{banerjee,prb,manosa,liu2}.

\par
An important observation of our work is the sharp metamagnetic jump in the low-$T$ $M(H)$ curve. Similar metamagnetic transitions have already been reported in case of other members of the series, namely, Er$_2$CoMnO$_6$, Tb$_2$CoMnO$_6$, Sm$_2$CoMnO$_6$. ~\cite{banerjee,kim,kim2}. For the present TCMO sample, the metamagnetism shows some unusual characteristics. (i) It is not present in the virgin leg [OA curve in fig. 3(a)]. (ii) The metamagnetic jump is absent at $T \geq$ 10 K. (iii) The hysteresis loops at 2 K and 5 K [see figs. 3(a-b)] have a elongated quadrilateral-like shape. 
\par
It has been already mentioned that the magnetic interactions $J_{Tm-Co}$, $J_{Tm-Mn} <$ 0 (AFM), while $J_{Co-Mn} >$ 0 (FM) ~\cite{kim,banerjee,tmcmo}. The Tm presumably attains an ordered magnetic state below about 8 K (inset of fig. 2 (a)).  Since, $J_{Tm-Co}$, $J_{Tm-Mn} <$ 0 (AFM), the Tm sublattice is likely to be antiparallel to the Co/Mn sublattices. The jump can occur due to the rotation of the Tm moment towards the spin direction of the ferromagnetically ordered Co/Mn sublattices.

\par
At 10 K, hysteresis loop is observed in th $M-H$ data with coercivity of 4.5 kOe [see fig. 3(c)], although $M$ does not saturate even at 70 kOe of field, and it almost increases linearly with $H$ beyond the loop. The 30 K isotherm also show small narrow hysteresis (coercivity of 0.5 kOe), and there is no signature of saturation [see fig. 3(d)]. This evidently contradicts the powder neutron diffraction result, which predicts a purely FM phase below 59 K due to the parallel arrangements of Co and Mn ions ~\cite{tmcmo}. Possibly, the magnetic structure between 30-59 K is not purely FM, and there is some degree of spin canting. It is to be noted that the neutron magnetic form factors of Co$^{2+}$ and Mn$^{4+}$ are very close ~\cite{zhelud}, and therefore, reflections due to a canted structure between Co$^{2+}$ and Mn$^{4+}$ may escape the observation in the previous neutron diffraction study.

\par
Most interestingly, the nature of the  $M(H)$ curve at 2 K is significantly different from the 10 K one [see figs. 3(a) and (c)] When the field is initially applied, Co and Mn moments get oriented easily towards magnetic field direction due to FM Co$^{2+}$-O$^{2-}$-Mn$^{4+}$ super-exchange interaction and weak anisotropy. On the other hand, Tm moments remain oppositely aligned to Co/Mn moment because of the AFM coupling. As the field is increased, the Zeeman energy starts to dominate over the exchange interaction. Thus on reaching a critical value of the magnetic field ($H_c$ = $\pm$32 kOe), sharp metamagnetic jump is observed and above this field the Tm-moments flipped. The $M$ value at 70 kOe ($\sim$ 10 $\mu_B/f.u.$) exceeds the expected value for a fully saturated Co/Mn sublattice (6 $\mu_B/f.u.$), indicating the partial contribution from the Tm moments.

\par 
It is to be noted that no metamagnetic jump is observed at 2 K in the virgin leg segment OA as marked in fig. 3(a), although a change in slope near the jump field. The metamagnetic jump is only observed when the field is decreased ($+$70 kOe to $-$70 kOe, curve ABC) or increased ($-$70 kOe to $+$70 kOe, curve CDA). There is no jump when the field is decreased/increased in a particular quadrant (say, AB in the first quadrant, or CD in the third quadrant). It appears that for jump to occur, the system should previously attain a field value higher than the metamagnetic field. It appears that the system gets magnetically aligned above $H_c$ in the direction of $H$, and a sufficient negative field is required to align it in the direction of $-H$, which occurs through the metamagnetic jump in the third quadrant (BC). Similar phenomenon occurs, when the system is exposed to a large negative field ($H < -H_c$), and a large positive field can only align it in the first quadrant though the jump (EF).  Our thermo-remanent magnetization also indicates that the application of large field tend to keep the system in a spin-aligned state even when the field is removed. Our measurement of $M$ in zero magnetic field while heating depends strongly on the history of cooling. A larger cooling field ($\geq$ 10 kOe), provides larger value of $M$ while heating at least up tp 40 K.

\par
In our heat capacity data, two distinct transitions are observed, {\it e.g.} ferromagnetic orientation of Co-Mn moments around $T_c$ = 60 K and another is around $T_s$ = 8 K. We have calculated the change in magnetic entropy (magneto-caloric effect, $\Delta S$) due to the application of magnetic field from the $C_p$ vs $T$ data measured at two different fields. We can notice also that the amount of entropy change at these two transitions points are very much comparable. These changes can be attributed to the suppression of spin fluctuations by $H$  around the transition points. Interestingly, we do not find any signature in the $\Delta S$ vs $T$ data around $T_N^{Tm}$ ($\sim$ 20 K). This is possibly due to the AFM arrangement of Tm and Co/Mn moments, which probably remains unaffected by the applied magnetic field. The low-$T$ peak of $\Delta S$ at around $T_s$ = 8 K, is possibly connected to the spin reorientation transition of the magnetic moments coming from Tm relative to Co, and Mn.   
\par 
There has been several reports on numbers of R$_2$CoMnO$_6$ (R= Lu, Y, Gd, Sm, Er, Tb) of exhibiting magneto-dielectric properties ~\cite{kim,moon1,moon2,kim2,tmcmo}. Even multiferroicity has been claimed for Lu$_2$CoMnO$_6$, Y$_2$CoMnO$_6$ arising out of symmetric exchange striction activated by broken inversion symmetry of up-up-down-down ($\uparrow \uparrow \downarrow \downarrow$) spin arrangement with alternating charge order. But for Tm$_2$CoMnO$_6$, no such detailed work has been done yet. In our work, we found a clear dielectric anomaly at the onset point of FM ordering temperature that was unnoticed in earlier report ~\cite{tmcmo}. The simultaneous occurrence of electric and magnetic orderings triggered the chances of magneto-electric coupling below the transition temperature. Moreover, we found real part of dielectric permittivity is very high ($\epsilon^{\prime}\sim$ 2700 ) that is advantageous for technological applications. As well as, such high dielectric constants is usual in non-cetrosymmetric structured materials, but for perovskites system this is quite uncommon. $\epsilon^{\prime}(T)$ becomes independent of $T$ below 200 K, whereas above that a sharp rise is occurred indicating towards some kind of relaxation behaviour. Such type of DP primarily shows Maxwell-Wagner type of relaxation which is may be due to the inhomogeneous grain-grain boundary effect, very common for polycrystalline oxides ~\cite{banerjee3,MW}. Besides that, the magnitude of activation energy and time constant found from the $T$ variation of $ln\tau$ (whereas, $\tau$ is obtained from $tan\delta$ vs $T$ plot) indicates towards some hopping mechanism where the charge carrier is heavier than normal electronic mass.

In conclusion, we observe several untold and noteworthy anomalous features in magnetic and dielectric behaviors of the double perovskite Tm$_2$CoMnO$_6$. The compound shows thermal hysteresis in $M(T)$ and metamagnetic jumps in low-$T$ isothermal magnetization in its AFM-like ground state. Anomalous thermal characteristics and sizable magnetocaloric response are obtained from the heat capacity study. It has a high dielectric constant with a indicative hopping conductive nature. Occurrence of structural and dielectric anomalies at transition temperature points towards possible multiferroicity that can be confirmed through polarization measurements. Moreover, the present work is executed on a polycrystalline sample with random orientation of anisotropy axis, but a detailed work on single crystalline sample is essential to obtain a comprehensive understanding of the system.  

\section{Acknowledgment}
A. Banerjee wishes to thank DST-SERB program for his NPDF fellowship (PDF/2021/004484). The Department of Science and Technology (India) is acknowledged for financial support, and the Saha Institute of Nuclear Physics and Jawaharlal Nehru Centre for Advanced Scientific Research, India for facilitating the experiments at the Indian Beam Line, Photon Factory, KEK, Japan. The work is supported through the financial grants from FIST project, DST, Government of India (Ref: SR/FST/PS-II/2018/52, TPN No-19862) and CAS Program, UGC, Government of India (No.F.530/20/CAS-II/2018 (SAP-I) Dt. 25.07.2018).


\begin{thebibliography}{99}
	
\bibitem{memory1} X. Chen, Y. Zhou, V. A. L. Roy and S.-T. Han, Adv. Mater. {\bf 30}, 1703950 (2018).

\bibitem{memory2} E. Carlos, R. Branquinho, R. Martins, A. Kiazadeh and E. Fortunato, Adv. Mater. {\bf 33}, 2004328 (2021).

\bibitem{medical} M. P. Nikolov and M. S. Chavali, Biomimetics (Basel) {\bf 5}, 27 (2020).

\bibitem{energy1} Z. Lei, J. M. Lee, G. Singh, C.I. Sathish, X. Chu, A. H. Al-Muhtaseb, A. Vinu and J. Yi, Energy Storage Materials {\bf 36}, 514-550 (2021)

\bibitem{energy2} C. Chappert, A. Fert, and F. Dau, Nat. Mater. {\bf 6}, 813 (2007).

\bibitem{remote} X. Yu, T. Marks and A. Facchetti, Nat. Mater. {\bf 15}, 383–396 (2016).

\bibitem{sharma} G. Sharma, J. Saha, S.D. Kaushik, V. Siruguri and S Patnaik, Appl. Phys. Lett. {\bf 103}, 012903 (2013).

\bibitem{chikara} S. Chikara, J. Singleton, J. Bowlan, D. A. Yarotski, N. Lee, H. Y. Choi, Y. J. Choi, and V. S. Zapf, Phys. Rev. B {\bf 93}, 180405 (2016).

\bibitem{terada} N. Terada, D. D. Khalyavin, P. Manuel, W. Yi, H. S. Suzuki, N. Tsujii, Y. Imanaka and A. A. Belik, Phys. Rev. B {\bf 91}, 104413 (2015). 

\bibitem{kim} J.H. Kim, K.W. Jeong, D.G. Oh, H.J. Shin, J.M. Hong, J.S. Kim, J.Y. Moon, N. Lee and Y.J. Choi, Scientific Reports {\bf 11}, 23786 (2021).

\bibitem{moon1} J. Y. Moon, M. K. Kim, D. G. Oh, J. H. Kim, H. J. Shin, Y. J. Choi and N. Lee, Phys. Rev. B {\bf 98}, 174424 (2018). 

\bibitem{moon2} J.Y. Moon, M.K. Kim, Y.J. Choi and N. Lee, Sci. Rep. {\bf 7}, 16099 (2017). 

\bibitem{balli} M. Balli, P. Fournier, S. Jandl, K.D. Truong and M.M. Gospodinov, J. Appl. Phys. {\bf 116}, 073907 (2014). 


\bibitem{liu} W. Liu, L. Shia, S. Zhou, J. Zhao, Y. Li and Y. Guo, J. Appl. Phys. {\bf 116}, 193901 (2014).

\bibitem{sfco} R. Pradheesh, H.S. Nair, V. Sankaranarayanan and K. Sethupathi, Appl. Phys. Lett. {\bf 101}, 142401 (2012). 

\bibitem{murthy} J. Krishna Murthy and A. Venimadhav, J. Alloys Compd. {\bf 719}, 341-346 (2017). 

\bibitem{metamag1} H. S. Nair, R. Pradheesh, Y. Xiao, D. Cherian, S. Elizabeth, T. Hansen, T. Chatterji and Th. Bruckel, J. Appl. Phys. {\bf 116}, 123907 (2014).

\bibitem{cao} G. Cao, T. F. Qi, L. Li, J. Terzic, S. J. Yuan, L. E. DeLong, G. Murthy and R. K. Kaul, Phys. Rev. Lett. {\bf 112}, 056402 (2014).

\bibitem{kim2} M. K. Kim, J. Y. Moon, S. H. Oh, D. G. Oh, Y. J. Choi and N. Lee,  Sci. Rep., {\bf 9}, 5456 (2019). 

\bibitem{ding} X. Ding, B. Gao, E. Krenkel, C. Dawson, J. C. Eckert, S.-W. Cheong and V. Zapf, Phys. Rev. B {\bf 99}, 014438 (2019).

\bibitem{kimura} T. Kimura, T. Goto, H. Shintani, K. Ishizaka, T. Arima and Y. Tokura, Nature {\bf 426}, 55–58 (2003).

\bibitem{tmcmo} J. Blasco, J. L. García-Muñoz, J. García, G. Subías, J. Stankiewicz, J. A. Rodríguez-Velamazán, and C. Ritter, Phys. Rev. B {\bf 96}, 024409 (2017).

\bibitem{kumar} S. Kumar, G. Giovannetti, J. van den Brink, and S. Picozzi, Phys. Rev. B {\bf 82}, 134429 (2010). 

\bibitem{veni} J. Krishna Murthy, K. D. Chandrasekhar, H. C.Wu, H. D. Yang, J. Y. Lin and A. Venimadhav, Eur. Phys. Lett. {\bf 108} 27013 (2014). 

\bibitem{tapan_chatt} T Chatterji, B. Frick, and H. S Nair, J. Phys.: Condens. Matter {\bf 24}, 266005 (2012).

\bibitem{kim3} M. K. Kim, J. Y. Moon, H. Y. Choi, S. H. Oh, N. Lee, and Y. J. Choi, J. Phys.: Condens. Matter {\bf 27}, 426002 (2015).

\bibitem{banerjee} A. Banerjee, J. Sannigrahi, S. Giri and S. Majumdar, Phys. Rev. B {\bf 98}, 104414 (2018)

\bibitem{disorder} A. J. Bar\'on-Gonz\'alez, C Frontera, J. L. Garc\'ia-Mu\~noz, B. Rivas-Murias and J. Blasco, J. Phys.: Condens. Matter {\bf 23}, 496003 (2011). 

\bibitem{maud} http://maud.radiographema.eu/.

\bibitem{cation_disorder1} A. S. Ogale, S. B. Ogale, R. Ramesh and T. Venkatesan, Appl. Phys. Lett. {\bf 75}, 537 (1999).

\bibitem{cation_disorder2} R. C. Sahoo, Y. Takeuchi, A. Ohtomo and Z. Hossain, Phys. Rev. B {\bf 100}, 214436 (2019).

\bibitem{cation_disorder3} C. Meneghini, Sugata Ray, F. Liscio, F. Bardelli, S. Mobilio and D. D. Sarma, Phys. Rev. Lett. {\bf 103}, 046403 (2009).

\bibitem{Gd-order} J. Krishna Murthy, K. D. Chandrasekhar, S. Mahana, D. Topwal and A. Venimadhav, J. Phys. D: Appl. Phys. {\bf 48}, 355001 (2015).

\bibitem{banerjee2} A. Banerjee, J. Sannigrahi, S. Giri and S. Majumdar, J. Phys.: Condens. Matter {\bf 29}, 115803 (2017).

\bibitem{siruguri} V. Siruguri, P. D. Babu, S. D. Kaushik, A. Biswas, S. K. Sarkar, M. Krishnan and P. Chaddah, J. Phys.: Condens. Matter {\bf 25}, 496011 (2013).

\bibitem{sbroy} S. B. Roy, M. K. Chattopadhyay, P. Chaddah and A. K. Nigam, Phys. Rev. B {\bf 71}, 174413 (2005).

\bibitem{chadda} A. Kumari, C. Dhanasekhar, P. Chaddah, D, Chandrasekhar Kakarla, H. D. Yang, Z H Yang, and B. H. Chen, Y. C. Chung, and A. K. Das, J. Phys.: Condens. Matter {\bf 32}, 155803 (2020).

\bibitem{nag} A. Nag, S. Bhowal, A. Chakraborty, M. M. Sala, A. Efimenko, F. Bert, P. K. Biswas, A. D. Hillier, M. Itoh, S. D. Kaushik, V. Siruguri, C. Meneghini, I. Dasgupta and Sugata Ray, Phys. Rev. B {\bf 98}, 014431 (2018).

\bibitem{mce} V. K. Pecharsky and K. A. Gschneidner, J. Appl. Phys. {\bf 86}, 565 (1999).

\bibitem{transport} R.C. Sahoo, S. Das and T.K. Nath, J. Magn. Magn. Mater. {\bf 60}, 409-417 (2018).

\bibitem{wang} L. Y. Wang, Q. Li, Y. Y. Gong, D. H. Wang, Q. Q. Cao and Y. W. Du, J. Am. Ceram. Soc. {\bf 97}, 2024–2026 (2014).

\bibitem{highEpsilon} C. C. Wang and L. W. Zhang, New J. Phys. {\bf 9}, 210 (2007).

\bibitem{banerjee3} A. Banerjee, J. Sannigrahi, S. Giri, and S. Majumdar, Phys. Status Solidi B {\bf 253}, 9 (2016).

\bibitem{yang} W.Z. Yang, X.Q. Liu, H.J. Zhao and X.M. Chen, J. Magn. Magn. Mater. {\bf 371}, 52–59 (2014).

\bibitem{lunk} P. Lunkenheimer, R. Fichtl, S. G. Ebbinghaus and A. Loidl, Phys. Rev. B {\bf 70}, 172102 (2004).

\bibitem{bid} O. Bidault, M. Maglione, M. Actis, M. Kchikech and B. Salce, Phys. Rev. B {\bf 52}, 4191 (1995).

\bibitem{camn7o12} J. Sannigrahi, S. Chattopadhyay, D. Dutta, S. Giri and S. Majumdar, J. Phys.: Condens. Matter {\bf 25}, 246001 (2013).

\bibitem{fsma} S. Chatterjee, S. Giri, S.K. De and S. Majumdar, Phys. Rev. B {\bf 79}, 092410 (2009).

\bibitem{fsma2} S. Pramanick, P. Dutta, S. Chatterjee, S. Giri and S. Majumdar, J. Alloy. Compd. {\bf 657}, 313-317 (2016).

\bibitem{prb} P. Dutta, M. Das, A. Banerjee, S. Chatterjee and S. Majumdar, J. Alloy. Compd. {\bf 886}, 161198 (2021).

\bibitem{manosa} A. Planes, L. Ma\~nosa and M. Acet, J. Phys.: Condens. Matter {\bf 21}, 233201 (2009).

\bibitem{liu2} E. Liu, W. Wang, L. Feng, W. Zhu, G. Li, J. Chen, H. Zhang, G. Wu, C. Jiang, H. Xu and F. de Boer, Nat. Comm. {\bf 3}, 873 (2012).

\bibitem{zhelud} http://neutron.ornl.gov/~zhelud/useful/formfac/index.html

\bibitem{MW} S. Lafuerza, J. García, G. Sub\'ias, J. Blasco, K. Conder, and E. Pomjakushina, Phys. Rev. B {\bf 88}, 085130 (2013).


\end{thebibliography}
\end{document}